\RequirePackage{lineno}
\documentclass[aps,preprint,tightenlines,superscriptaddress,showpacs,byrevtex]{revtex4-1}

\usepackage{graphicx}
\usepackage{dcolumn}
\usepackage{bm}
\usepackage{rotating}
\usepackage{epstopdf}
\usepackage{color}
\usepackage{verbatim} 
\usepackage{multirow}
\usepackage[abs]{overpic}
\usepackage{amsmath}

\newcommand{\PreserveBackslash}[1]{\let\temp=\\#1\let\\=\temp}
\newcolumntype{C}[1]{>{\PreserveBackslash\centering}p{#1}}
\newcolumntype{R}[1]{>{\PreserveBackslash\raggedleft}p{#1}}
\newcolumntype{L}[1]{>{\PreserveBackslash\raggedright}p{#1}}

\newcommand{\EE}{e^+e^-}

\newcommand{\too}{\rightarrow}

\begin{document}
\graphicspath{{figure/}}
\DeclareGraphicsExtensions{.eps,.png,.ps}

\title{\quad\\[0.0cm] \boldmath Combined fit to the cross sections of $\EE \too \pi Z_c(3900) \too \pi\pi J/\psi$ and $\EE \too \pi Z_c(3900) \too \pi D\bar{D}^{*}$}

\author{Jielei Zhang}
\email{zhangjielei@ihep.ac.cn}
\author{Tengjiao Zhu}
\affiliation{College of Physics and Electronic Engineering, Xinyang Normal University, Xinyang 464000, People's Republic of China}

\begin{abstract}
The cross sections of $\EE \too \pi Z_c(3900) \too \pi\pi J/\psi$ and $\EE \too \pi Z_c(3900) \too \pi D\bar{D}^{*}$ have been measured by BESIII experiment.  We try to perform a combined fit to the cross sections with one Breit-Wigner function, the fit results show the structure's mass and width are $M=(4232\pm5)$ MeV/$c^2$, $\Gamma=(65\pm21)$ MeV. The ratio $\frac{\mathcal{B}(Z_c(3900)\too D\bar{D}^{*})}{\mathcal{B}(Z_c(3900)\too\pi J/\psi)}$ is determined to be $(16\pm6)$. We also try to fit the cross sections with two Breit-Wigner functions, while we can't come to any definitive conclusions about the second structure. More measurements are desired to improve the understanding of $\EE \too \pi Z_c(3900)$ line shape.
\end{abstract}

\maketitle
In recent years, charmonium physics gained renewed strong interest from both the theoretical and the experimental side, due to the observation of charmonium-like states, such as $X(3872)$~\cite{X3872}, $Y(4260)$~\cite{Y4260}, $Y(4360)$~\cite{Y4360} and $Y(4660)$~\cite{Y4660}. These states do not fit in the conventional charmonium spectroscopy, and could be exotic states that lie outside the quark model~\cite{theory1, theory11, theory22, theory33}. Moreover, in 2013, a charged charmonium-like state, $Z_c(3900)^{\pm}$, was observed by the BESIII~\cite{Zc1} and Belle Collaborations~\cite{Zc2} in the $\pi^{\pm}J/\psi$ invariant mass distribution of the process $e^+e^-\too\pi^+\pi^-J/\psi$ and confirmed using CLEO-c's data~\cite{Zc3}. As there are at least four quarks in the structure, many theoretical interpretations of the nature and the decay dynamics of the $Z_c(3900)$ have been put forward~\cite{theory2, theory3}.

Shortly afterwards, a similar charged structure, the $Z_c(3885)^{\pm}$, was observed in the $(D\bar{D}^{*})^{\pm}$ invariant mass distribution of the process $\EE \too \pi^{\pm}(D\bar{D}^{*})^{\mp}$~\cite{Zc4, Zc5}, with a spin parity ($J^P$) assignment of $1^+$ favored over the $1^-$ and $0^-$ hypotheses. And its mass and width are consistent with those of the $Z_c(3900)^{\pm}$. Recently, the spin and parity of the $Z_c(3900)^{\pm}$ are determined to be $J^P=1^+$~\cite{Zc6}. So now $Z_c(3900)^{\pm}$ and $Z_c(3885)^{\pm}$ have been taken as one state $Z_c(3900)$ in PDG~\cite{pdg}. The neutral structure, $Z_c(3900)^{0}$, also has been observed in the $\pi^{0}J/\psi$ invariant mass distribution of the process $\EE \too \pi^0\pi^0J/\psi$~\cite{Zc7, Zc8} and $(D\bar{D}^{*})^{0}$ invariant mass distribution of the process $\EE \too \pi^{0}(D\bar{D}^{*})^{0}$~\cite{Zc9}.

Until now, the nature of $Z_c(3900)$ is still unclear. Although $Z_c(3900)$ has charge indicates it is not conventional meson consisting of a quark-antiquark pair, its exact quark configuration is still unknown. Several models have been developed to describe its inner structure, including loosely bound hadronic molecules of two charmed mesons, compact tetraquarks, and hadro-quarkonium~\cite{theory2, theory3}. In this paper, we try to analyze all the cross sections of $\EE \too \pi Z_c(3900)$ to get the combined results about the $Z_c(3900)$. It maybe useful to understand the nature of $Z_c(3900)$.

\begin{figure}[htbp]
\begin{center}
\begin{overpic}[width=0.45\textwidth]{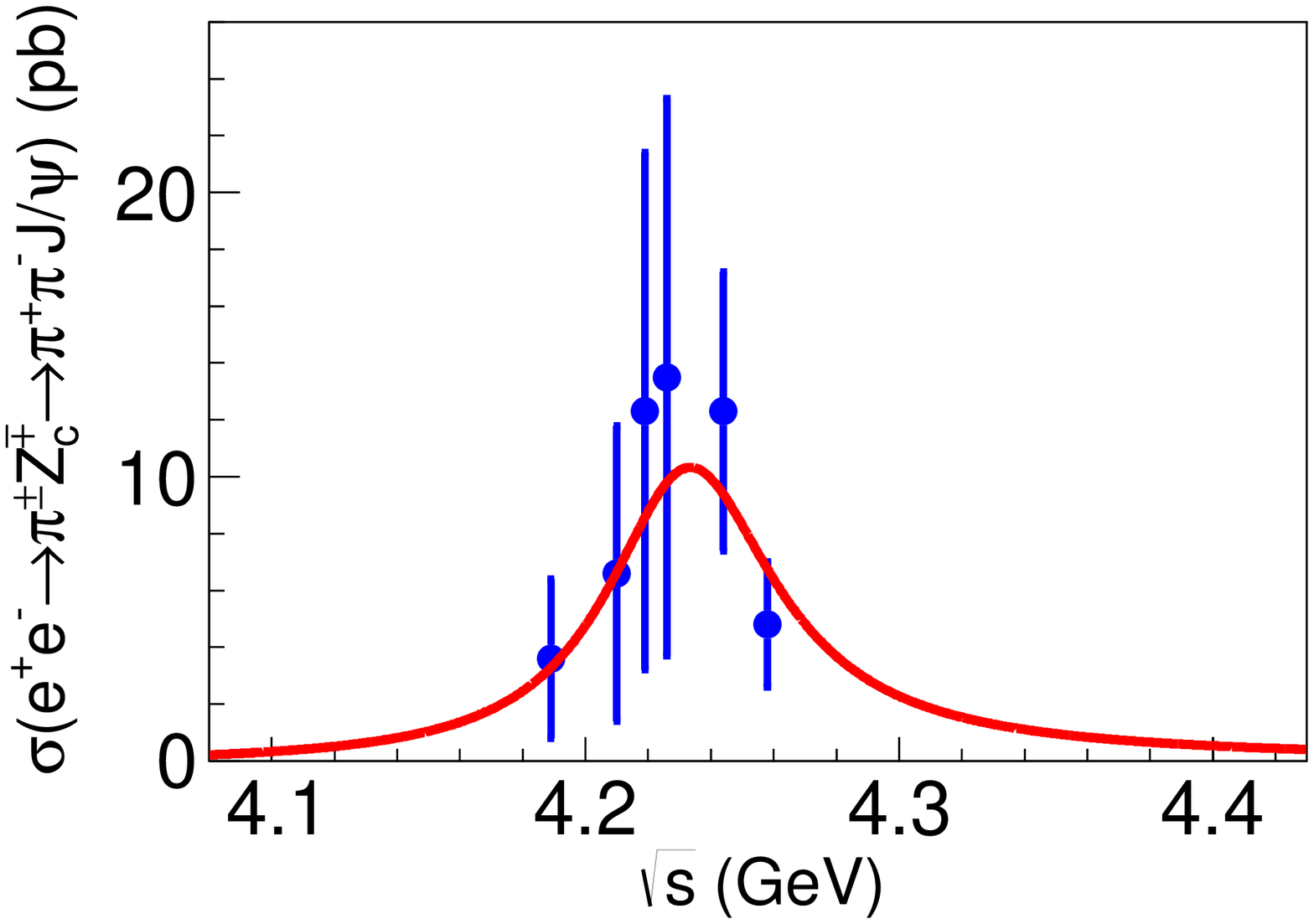}
\put(178,126){\large (a)}
\end{overpic}
\begin{overpic}[width=0.45\textwidth]{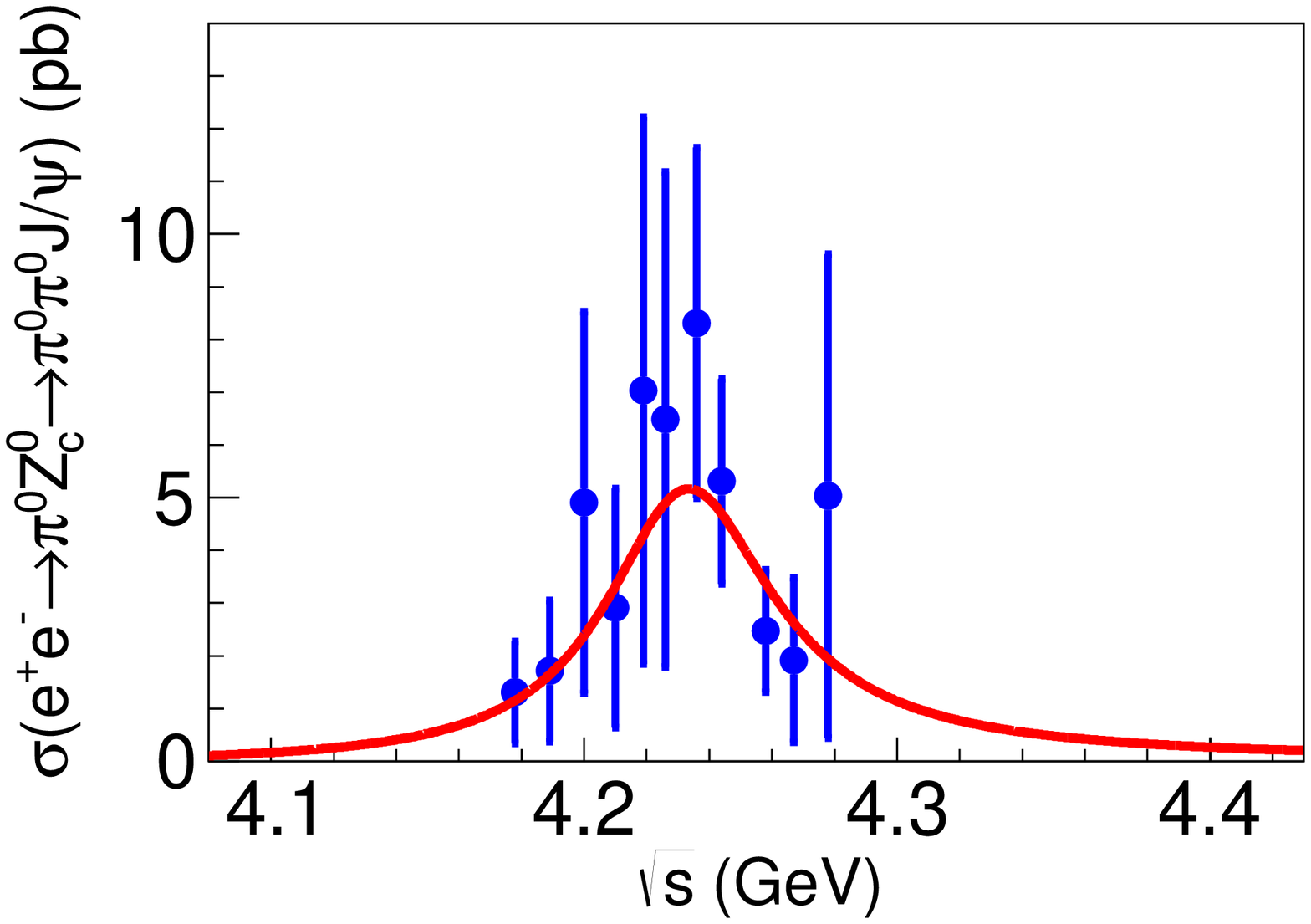}
\put(178,126){\large (b)}
\end{overpic}
\begin{overpic}[width=0.45\textwidth]{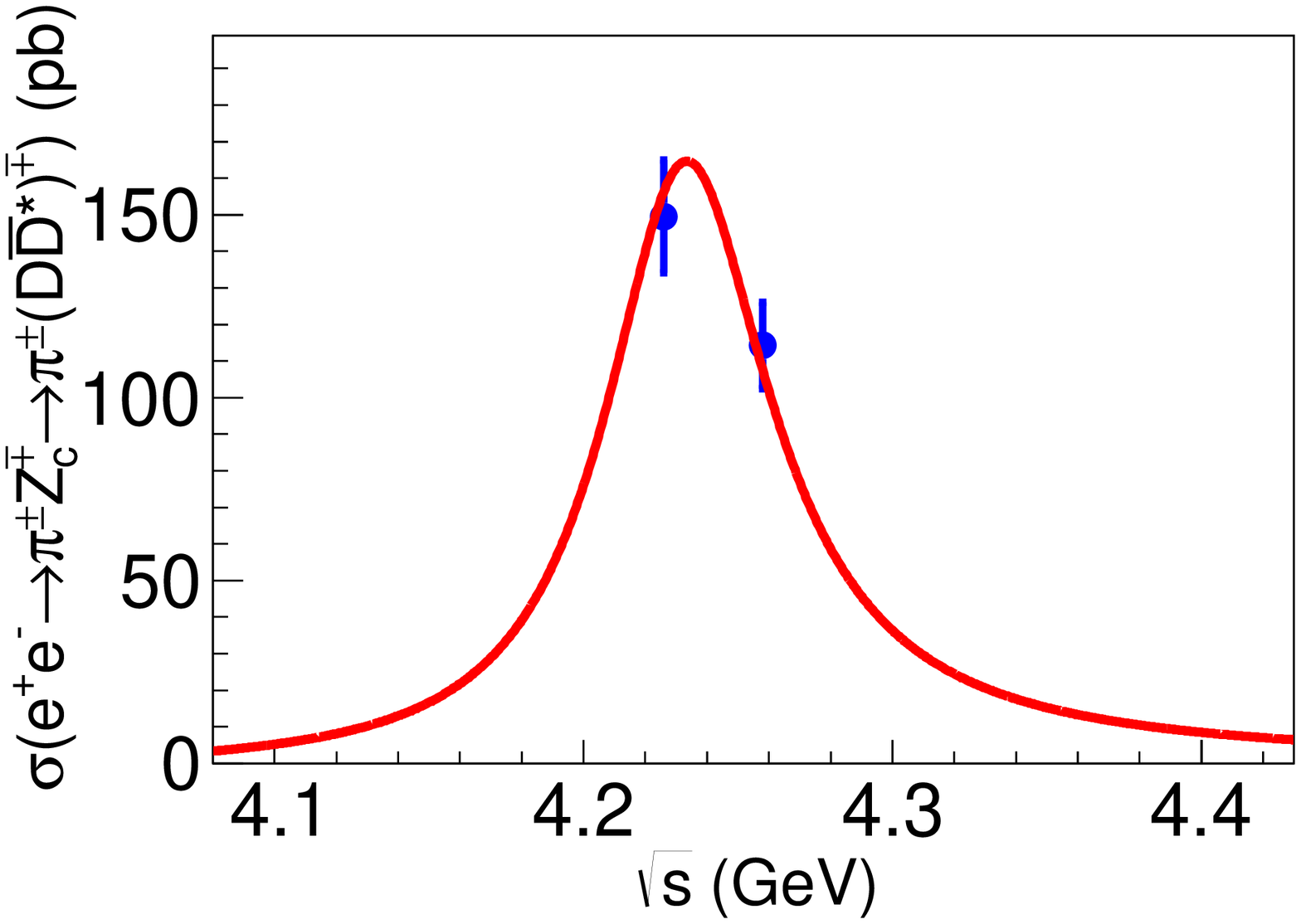}
\put(178,126){\large (c)}
\end{overpic}
\begin{overpic}[width=0.45\textwidth]{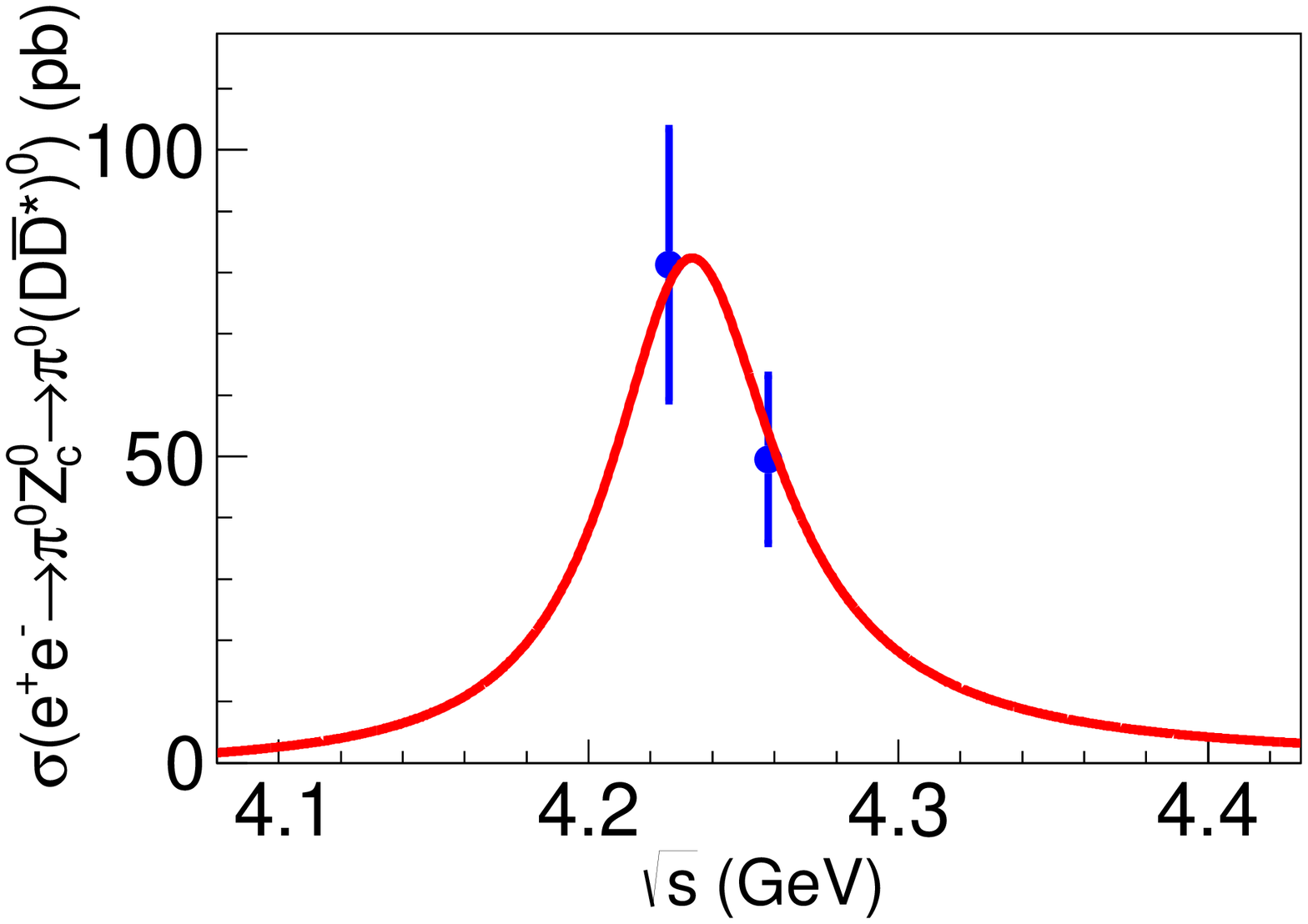}
\put(178,126){\large (d)}
\end{overpic}
\caption{The results of the combined fit to the cross sections of $\EE \too \pi^{\pm}Z_c(3900)^{\mp} \too \pi^+\pi^-J/\psi$ (a), $\EE \too \pi^{0}Z_c(3900)^{0} \too \pi^0\pi^0J/\psi$ (b), $\EE \too \pi^{\pm}Z_c(3900)^{\mp} \too \pi^{\pm}(D\bar{D}^{*})^{\mp}$ (c), and $\EE \too \pi^{0}Z_c(3900)^{0} \too \pi^{0}(D\bar{D}^{*})^{0}$ (d) using one Breit-Wigner function. The solid red curves show the best fits.}
\label{fig:crosssectionfit1}
\end{center}
\end{figure}

Figure~\ref{fig:crosssectionfit1} shows the cross sections of $\EE \too \pi^{0}Z_c(3900)^{0} \too \pi^0\pi^0J/\psi$, $\EE \too \pi^{\pm}Z_c(3900)^{\mp} \too \pi^{\pm}(D\bar{D}^{*})^{\mp}$ and $\EE \too \pi^{0}Z_c(3900)^{0} \too \pi^{0}(D\bar{D}^{*})^{0}$ below 4.3 GeV measured by BESIII experiment~\cite{Zc5, Zc8, Zc9}, here the error bars represent the total errors. It should be noted that the cross sections rely on the model for the $Z_c$ lineshapes used in the experiment. The cross sections of $\EE \too \pi^{\pm}Z_c(3900)^{\mp} \too \pi^+\pi^-J/\psi$ in Fig.~\ref{fig:crosssectionfit1} are obtained by the results of cross sections of $\EE \too \pi^{0}Z_c(3900)^{0} \too \pi^0\pi^0J/\psi$~\cite{Zc8} and cross sections of $\EE \too \pi^+\pi^-J/\psi$~\cite{Zc10} assuming the isospin conservation. We can see the line shapes are similar, the cross sections around 4.23 GeV are larger than those around 4.26 GeV, which indicates that one Breit-Wigner function with mass around 4.23 GeV can describe the cross sections. Assuming $\pi Z_c(3900)$ comes from one resonance, we perform a combined fit to the cross sections using a least $\chi^{2}$ method. We fit the cross sections with one constant width relativistic Breit-Wigner function, that is,
\begin{equation}
\begin{aligned}
\sigma(\sqrt{s})= & |BW(\sqrt{s})\sqrt{\frac{PS(\sqrt{s})}{PS(M)}}|^{2},
\end{aligned}
\end{equation}
where $PS(\sqrt{s})$ is the two-body phase space factor that increases smoothly from the mass threshold with the $\sqrt{s}$~\cite{pdg}, $BW(\sqrt{s})=\frac{\sqrt{12\pi\Gamma_{ee}\mathcal{B}(\pi Z_c(3900))\Gamma}}{s-M^{2}+iM\Gamma}$, is the Breit-Wigner function for a vector state, with mass $M$, total width $\Gamma$, electron partial width $\Gamma_{ee}$, and the branching fraction to $\pi Z_c(3900)$, $\mathcal{B}(\pi Z_c(3900))$. From the fit, the $\Gamma_{ee}$ and $\mathcal{B}(\pi Z_c(3900))$ can not be obtained separately, we can only extract the product $\Gamma_{ee}\mathcal{B}(\pi Z_c(3900))$.

In the combined fit, we fix the ratio $\frac{\mathcal{B}(\pi^{+}Z_c(3900)^{-})+\mathcal{B}(\pi^{-}Z_c(3900)^{+})}{\mathcal{B}(\pi^{0}Z_c(3900)^{0})}$ to be 2 assuming the isospin conservation. The $\chi^{2}$ is minimized to obtain the best estimation of the resonant parameters, and the uncertainties are obtained when $\chi^{2}$ value change 1 compared with the minimum value. Figure~\ref{fig:crosssectionfit1} shows the fit results. The fits indicate the existence of one resonance (called $Y$) with mass and width are $M=(4232\pm5)$ MeV/$c^{2}$ and $\Gamma=(65\pm21)$ MeV. The goodness of the fit is $\chi^{2}/\text{ndf}=5.3/17$, where $\text{ndf}$ is the number of degrees of freedom. The all fitted parameters of the cross sections are listed in Table~\ref{tab:fitresult1}.

\begin{table}[htbp]
\begin{center}
\caption{ The fitted parameters from the combined fit to the cross sections of $\EE \too \pi Z_c(3900) \too \pi\pi J/\psi$ and $\EE \too \pi Z_c(3900) \too \pi D\bar{D}^{*}$ using one Breit-Wigner function. Here, $\mathcal{B}(Y \too \pi Z_c(3900) \too \pi\pi J/\psi)$ means the sum of $\mathcal{B}(Y \too \pi^{+}Z_c(3900)^{-} \too \pi^{+}\pi^{-}J/\psi)$, $\mathcal{B}(Y \too \pi^{-}Z_c(3900)^{+} \too \pi^{+}\pi^{-}J/\psi)$ and $\mathcal{B}(Y \too \pi^{0}Z_c(3900)^{0} \too \pi^{0}\pi^{0}J/\psi)$, and the same definition for $\mathcal{B}(Y \too \pi Z_c(3900) \too \pi D\bar{D}^{*})$.  }
\label{tab:fitresult1}
\begin{tabular}{cc}
  \hline
  \hline
  \quad \quad Parameter \quad \quad & \quad \quad Solution \quad \quad  \\
  \hline
  \quad \quad $M$ (MeV/$c^{2}$) \quad \quad & \quad \quad $4232\pm5$ \quad \quad  \\
  \quad \quad $\Gamma$ (MeV) \quad \quad & \quad \quad $65\pm21$ \quad \quad  \\
  \quad \quad $\Gamma_{ee}\mathcal{B}(Y \too \pi Z_c(3900) \too \pi\pi J/\psi)$ (eV) \quad \quad & \quad \quad $1.2\pm0.3$ \quad \quad  \\
  \quad \quad $\Gamma_{ee}\mathcal{B}(Y \too \pi Z_c(3900) \too \pi D\bar{D}^{*})$ (eV) \quad \quad & \quad \quad $19.6\pm4.5$ \quad \quad  \\
  \hline
  \hline
\end{tabular}
\end{center}
\end{table}

From Table~\ref{tab:fitresult1}, the ratio $\frac{\mathcal{B}(Z_c(3900)\too D\bar{D}^{*})}{\mathcal{B}(Z_c(3900)\too\pi J/\psi)}$ is determined to be $(16\pm6)$, which shows the branching fraction of $Z_c(3900)$ decays into open charm final states is about an order of magnitude larger than that of $Z_c(3900)$ decays into hidden charm final states. It indicates that $Z_c(3900)$ has stronger coupling with open charm channel, which is important to understand the nature of $Z_c(3900)$.

\begin{figure}[htbp]
\begin{center}
\begin{overpic}[width=0.45\textwidth]{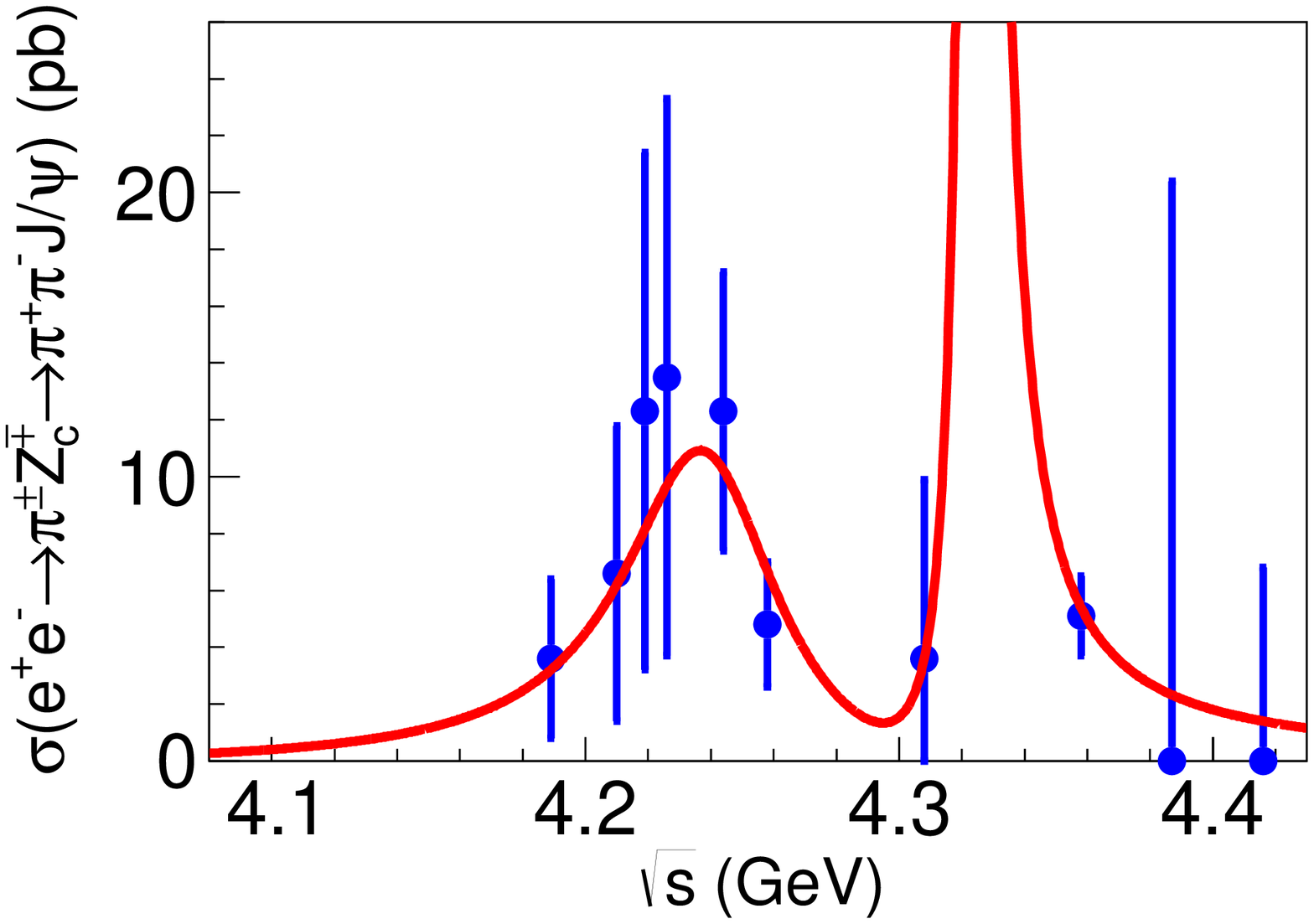}
\put(178,126){\large (a)}
\end{overpic}
\begin{overpic}[width=0.45\textwidth]{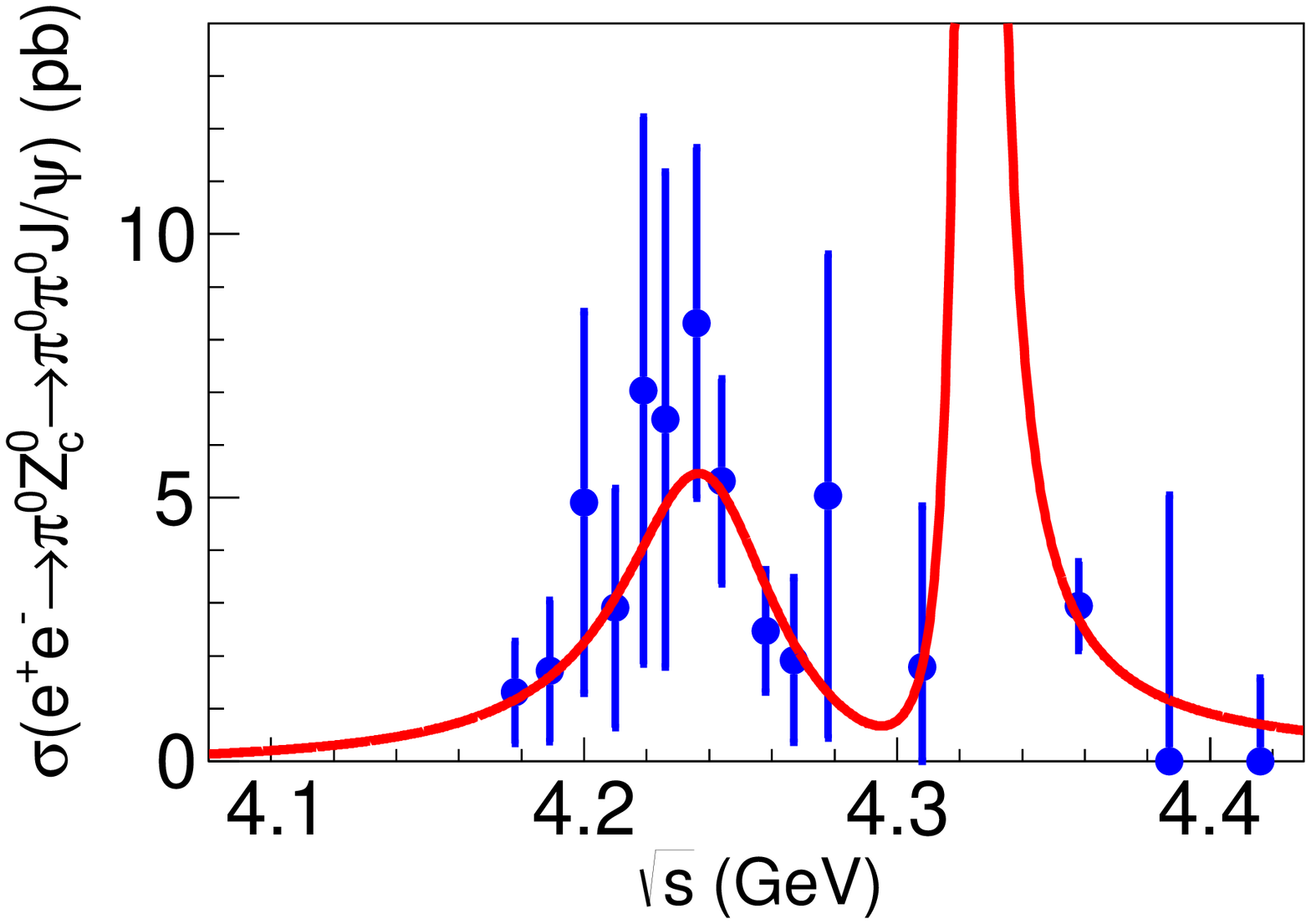}
\put(178,126){\large (b)}
\end{overpic}
\begin{overpic}[width=0.45\textwidth]{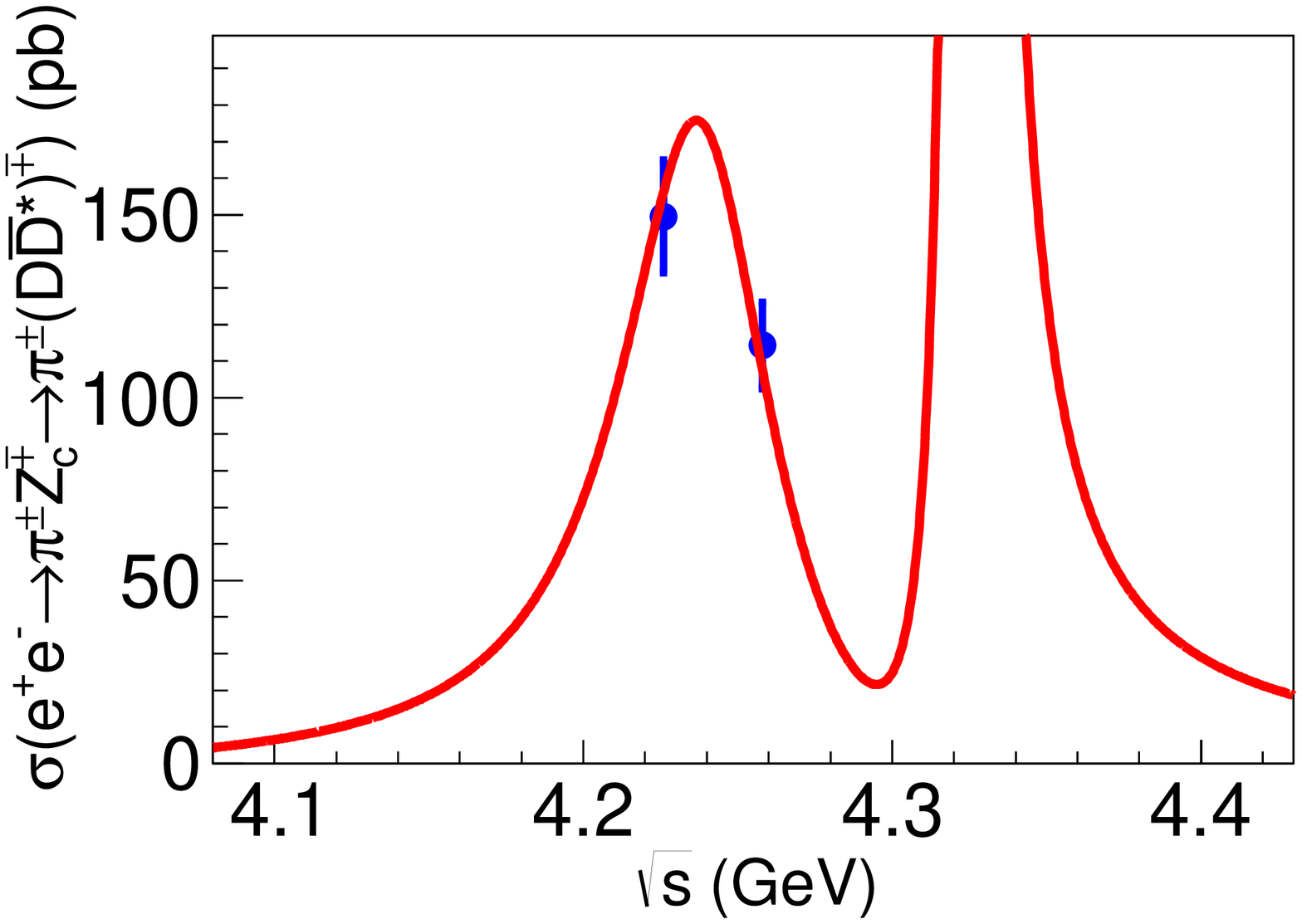}
\put(178,126){\large (c)}
\end{overpic}
\begin{overpic}[width=0.45\textwidth]{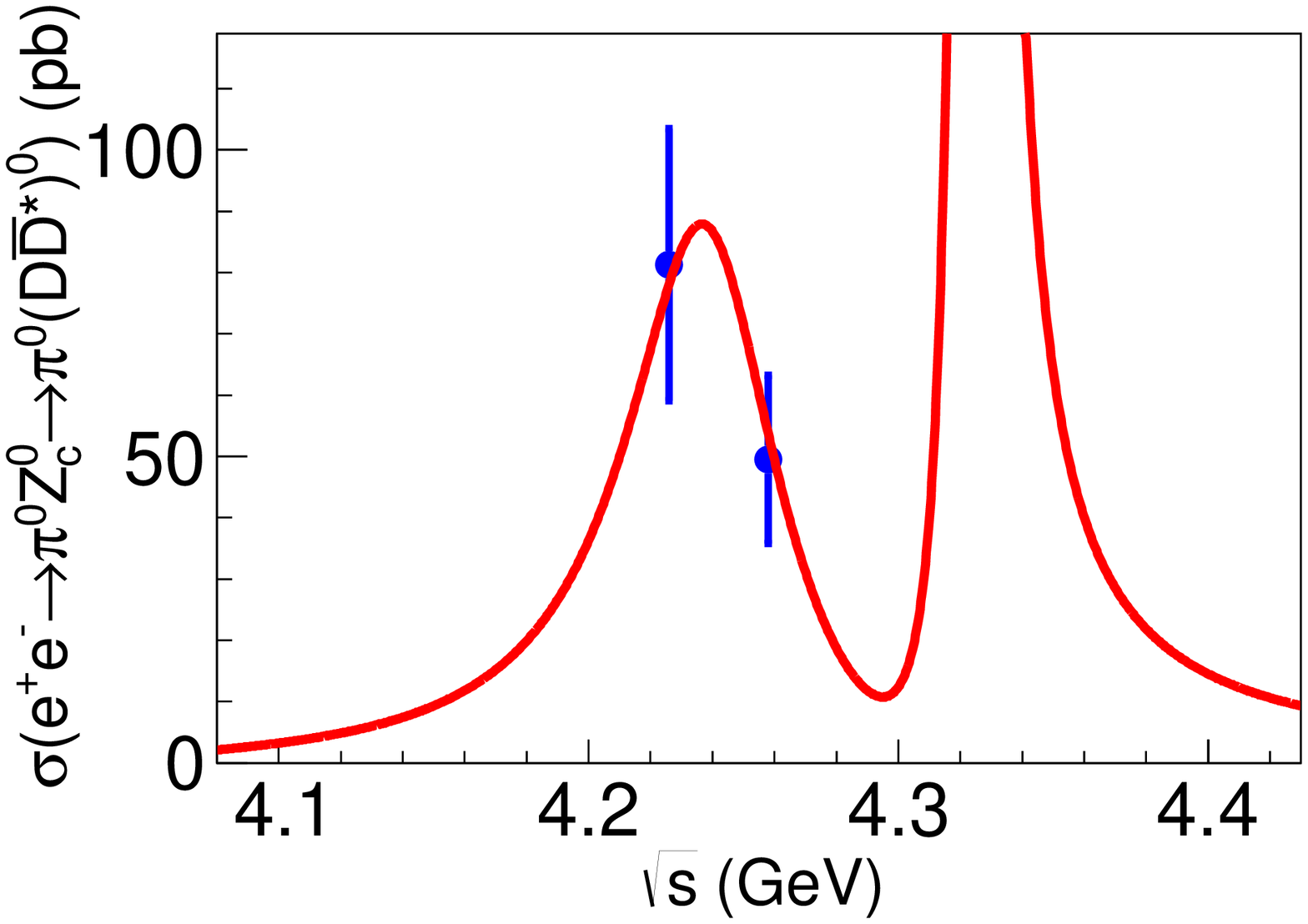}
\put(178,126){\large (d)}
\end{overpic}
\caption{The results of the combined fit to the cross sections of $\EE \too \pi^{\pm}Z_c(3900)^{\mp} \too \pi^+\pi^-J/\psi$ (a), $\EE \too \pi^{0}Z_c(3900)^{0} \too \pi^0\pi^0J/\psi$ (b), $\EE \too \pi^{\pm}Z_c(3900)^{\mp} \too \pi^{\pm}(D\bar{D}^{*})^{\mp}$ (c), and $\EE \too \pi^{0}Z_c(3900)^{0} \too \pi^{0}(D\bar{D}^{*})^{0}$ (d) using two Breit-Wigner functions. The solid red curves show the best fits.}
\label{fig:crosssectionfit2}
\end{center}
\end{figure}

Figure~\ref{fig:crosssectionfit2} shows the cross sections with some points above 4.3 GeV added~\cite{Zc7}. As shown by the fit results in Ref.~\cite{Zc10}, we also try to use coherent sum of two Breit-Wigner functions (called $Y_1$ and $Y_2$) to fit the cross sections assuming the isospin conservation. In the fit, $\mathcal{B}(Y_2 \too \pi Z_c(3900) \too \pi D\bar{D}^{*})$ is fixed to be $\frac{\mathcal{B}(Y_1 \too \pi Z_c(3900) \too \pi D\bar{D}^{*})\mathcal{B}(Y_2 \too \pi Z_c(3900) \too \pi\pi J/\psi)}{\mathcal{B}(Y_1 \too \pi Z_c(3900) \too \pi\pi J/\psi)}$. Figure~\ref{fig:crosssectionfit2} shows the fit results. The two structures' mass and width are $M_1=(4239\pm10)$ MeV/$c^{2}$, $\Gamma_1=(60\pm21)$ MeV and $M_2=(4325\pm16)$ MeV/$c^{2}$, $\Gamma_2=(8\pm26)$ MeV. Two solutions are found with the same fit quality, the all fitted parameters of the cross sections are listed in Table~\ref{tab:fitresult2}. The ratio $\frac{\mathcal{B}(Z_c(3900)\too D\bar{D}^{*})}{\mathcal{B}(Z_c(3900)\too\pi J/\psi)}$ is determined to be $(16\pm11)$ or $(16\pm7)$, which is consistent with the results of using one Breit-Wigner function to fit. Because the measured cross section points between 4.3 GeV and 4.4 GeV are very scarce, we can't come to some definitive conclusions about $Y_2$. It is clear that the result on the second peak in Fig.~\ref{fig:crosssectionfit2} can not be trusted, since the fit prefers a narrow structure in a region where no data sit. While the results of $Y_1$ are consistent with the above results of fitting with one Breit-Wigner function.

\begin{table}[htbp]
\begin{center}
\caption{ The fitted parameters from the combined fit to the cross sections of $\EE \too \pi Z_c(3900) \too \pi\pi J/\psi$ and $\EE \too \pi Z_c(3900) \too \pi D\bar{D}^{*}$ using two Breit-Wigner functions. }
\label{tab:fitresult2}
\begin{tabular}{ccc}
  \hline
  \hline
  \quad \quad Parameter \quad \quad & \quad \quad Solution I \quad \quad & \quad \quad Solution II \quad \quad  \\
  \hline
  $M_1$ (MeV/$c^{2}$) & \multicolumn{2}{c}{$4239\pm10$} \\
  $\Gamma_1$ (MeV) & \multicolumn{2}{c}{$60\pm21$}  \\
  $M_2$ (MeV/$c^{2}$) & \multicolumn{2}{c}{$4325\pm16$}  \\
  $\Gamma_2$ (MeV) & \multicolumn{2}{c}{$8\pm26$}  \\
  \quad \quad $\Gamma_{ee}\mathcal{B}(Y_1 \too \pi Z_c(3900) \too \pi\pi J/\psi)$ (eV) \quad \quad & \quad \quad $1.1\pm0.5$ \quad \quad & \quad \quad $1.6\pm0.5$ \quad \quad  \\
  \quad \quad $\Gamma_{ee}\mathcal{B}(Y_1 \too \pi Z_c(3900) \too \pi D\bar{D}^{*})$ (eV) \quad \quad & \quad \quad $18.0\pm9.1$ \quad \quad & \quad \quad $26.0\pm7.0$ \quad \quad  \\
  \quad \quad $\Gamma_{ee}\mathcal{B}(Y_2 \too \pi Z_c(3900) \too \pi\pi J/\psi)$ (eV) \quad \quad & \quad \quad $2.1\pm5.3$ \quad \quad & \quad \quad $2.6\pm6.3$ \quad \quad  \\
  \quad \quad $\phi$ (rad) \quad \quad & \quad \quad $0.1\pm1.5$ \quad \quad & \quad \quad $-1.3\pm1.2$ \quad \quad  \\
  \hline
  \hline
\end{tabular}
\end{center}
\end{table}

In summary, we perform a combined fit to the cross sections of $\EE \too \pi Z_c(3900) \too \pi\pi J/\psi$ and $\EE \too \pi Z_c(3900) \too \pi D\bar{D}^{*}$ with one Breit-Wigner function. The fit results show $\pi Z_c(3900)$ maybe from a structure with mass and width are $M=(4232\pm5)$ MeV/$c^2$, $\Gamma=(65\pm21)$ MeV. The ratio $\frac{\mathcal{B}(Z_c(3900)\too D\bar{D}^{*})}{\mathcal{B}(Z_c(3900)\too\pi J/\psi)}$ is determined to be $(16\pm6)$. We also try to use two Breit-Wigner functions to fit the cross sections, the first structure's results are consistent with the results of fitting with one Breit-Wigner function. Due to the measured cross section points between 4.3 GeV and 4.4 GeV are very scarce, we can't come to some definitive conclusions about the second structure. So the two resonances fit is nonphysical, and the one resonance fit should be taken as the result of the paper. At present, the number of data points in $\EE \too \pi Z_c(3900)$ line shape is relatively small and the errors are large, so more accurate measurements are desired to provide more constraint on the $\EE \too \pi Z_c(3900)$ line shape. It will help to understand where $Z_c(3900)$ comes from, and further reveal the nature of $Z_c(3900)$. These measurements can be achieved by BESIII and BelleII experiments in the further.

\section*{Acknowledgement}
This work is supported by the National Natural Science Foundation of China (NSFC) under Contract No. 11905179, Foundation of Henan Educational Committee (No. 19A140015), and Nanhu Scholars Program for Young Scholars of Xinyang Normal University.

\end{document}